# Electrodynamics of Photonic Temporal Interfaces

E. Galiffi[1,*], D. M. Solís[2], S. Yin[1,3], N. Engheta[4] and A. Alù[1,3,5,*]

[1] Photonics Initiative, Advanced Science Research Center, City University of New York, New York, USA

[2] Departamento de Tecnología de los Computadores y de las Comunicaciones, University of Extremadura, 10003 Cáceres, Spain

[3] Department of Electrical Engineering, City College of The City University of New York, New York, NY, 10031 USA

[4] Department of Electrical and Systems Engineering, University of Pennsylvania, Philadelphia, PA, 19104, USA

[5] Physics Program, Graduate Center of the City University of New York, New York, NY, 10016, USA2

*egaliffi@gc.cuny.edu
*aalu@gc.cuny.edu

**Abstract**

Exotic forms of wave control have been emerging by engineering matter in space and time. In this framework, temporal photonic interfaces, i.e., abrupt changes in the electromagnetic properties of a material, have been shown to induce temporal scattering phenomena dual to spatial reflection and refraction, at the basis of photonic time crystals and space-time metamaterials. Despite decades-old theoretical studies on these topics, and recent experimental demonstrations, the careful modeling of these phenomena has been lagging behind. Here, we develop from first principles a rigorous model of the electrodynamics of temporal photonic interfaces, highlighting the crucial role of the mechanisms driving time variations. We demonstrate that the boundary conditions and conservation laws associated with temporal scattering may substantially deviate from those commonly employed in the literature, based on their microscopic implementation. Our results open new vistas for both fundamental investigations over light-matter interactions in time-varying structures and for the prospect of their future implementations and applications in optics and photonics.

**Introduction**

Ultrafast temporal control of materials and metamaterials has recently emerged as a powerful tool for wave manipulation and control. In addition to serving as a fourth dimension beyond the three spatial ones, time empowers wave physics with a plethora of opportunities, enabling negative refraction[1,2], amplification, both based on photon pair generation[3-6] and cascaded photon upconversion[7,8], broken time-reversal symmetry and topological wave phenomena[9,10], synthetic optical drag[11] and nonreciprocity[12-17], photonic refrigeration[18] and thermal radiation engineering[19-20], as well as synthetic collisions between photons[21], emission control[22], filtering paradigms for classical[23] and quantum states of light[24], and pulse routing[25], among others[26]. At the basis of these phenomena are scattering phenomena dual to those enabled by spatial inhomogeneities. Over the years, it has been recognized that scattering from temporal boundaries is fundamentally distinct from spatial scattering, due to different causality relations and underlying conservation laws.

The foundational phenomena at the basis of temporal scattering are time-refraction - whereby a pulse is uniformly frequency-shifted as it experiences a temporal change in the optical properties of the host medium[27-28]; time-diffraction - whereby frequency fringes are produced upon scattering off an inhomogeneous structure whose electromagnetic response is varying in time at a rate comparable to a single optical cycle[29]; and time-reflection – consisting in broadband time-reversal of the wave generated by an abrupt, spatially homogeneous change in the wave impedance of the host material, forming a time-interface (TI)[30-33]. Carefully predicting the scattering products of time variations requires an understanding of the conservation laws that govern these non-conservative electrodynamic processes, and the corresponding boundary conditions. While this problem has a long history of theoretical investigations[30], its recent experimental demonstrations spurred renewed interest in exploring their accurate modeling[32]. In the following, we derive from first principles the boundary conditions emerging at a TI and highlight the role of the microscopic implementation of TIs in determining the quantities that are conserved in the process. These findings have implications on the energy required to drive temporal scattering events, and on the wave phenomena emerging from them.

**Results**

Maxwell's equations in the time-domain enforce the following relations among the instantaneous electromagnetic fields and the total current $\mathbf{J} = \mathbf{J}_c + \mathbf{J}_s$, formed by conduction $\mathbf{J}_c$ and impressed current $\mathbf{J}_s$,

$$\begin{aligned} \nabla \times \mathbf{E} &= -\partial_t \mathbf{B} \\ \nabla \times \mathbf{H} &= \partial_t \mathbf{D} + \mathbf{J} \end{aligned}, \quad (1)$$

where $\mathbf{E}$, $\mathbf{H}$ are the electric and magnetic fields, and $\mathbf{B}$, $\mathbf{D}$ are the magnetic induction and electric displacement. The common assumption in the broad literature on TIs is that, as the material properties abruptly change, and in the absence of material dispersion or currents ($\mathbf{J} = 0$), the magnetic induction $\mathbf{B}$ and the electric displacement $\mathbf{D}$ are conserved. The common justification for this assumption is that their continuity avoids the divergence of the time-derivatives in Eq. (1), and is associated with momentum conservation due to preserved spatial symmetry[30,34-35].

As we show in the following, and consistent with recent experiments[32], a TI does not necessarily conserve $\mathbf{B}$ and $\mathbf{D}$, but rather the boundary conditions strongly depend on the way in which the electromagnetic properties of the medium are changed in time. For instance, in the recent experimental demonstration of time reflections at a TI in a loaded transmission-line metamaterial[32], it was observed that, in certain configurations, rather than $\mathbf{B}$ and $\mathbf{D}$ the conserved quantities were $\mathbf{E}$ and $\mathbf{H}$. As a by-product, the local charge, proportional to $\mathbf{D}$, and the total momentum density $\mathbf{D} \times \mathbf{B}$ are not necessarily conserved at a TI, as commonly assumed in the literature. In order to debunk these results and understand the underlying conditions that determine the conservation law at TIs, in the following we develop a general model for TIs from first principles, within a framework dual to spatial interfaces, to derive the proper boundary conditions as a function of the underlying TI implementation.

Consider the boundary between two uniform transmission lines (Fig. 1a) or between two non-dispersive media (Fig. 1b). At position $z_0$ the material permittivity, or the capacitance per unit length of the transmission line, changes from $\varepsilon_1$ to $\varepsilon_2$, forming a spatial interface. The boundary conditions for the in-plane electric and magnetic fields can be derived by integrating Eq. (1) over a volume that contains the interface (as sketched in Fig. 1b) and using Gauss's theorem. Letting the volume vanish yields the conventional boundary conditions (see, e.g., Ref. 36)

$$\begin{aligned} E_x(z_0^+) - E_x(z_0^-) &= 0 \\ H_y(z_0^+) - H_y(z_0^-) &= J_s \end{aligned}, \quad (2)$$

where $J_s$ is the electric surface current density at the interface (in the absence of magnetic currents). For $J_s = 0$, the expected continuity of tangential $\mathbf{E}$ and $\mathbf{H}$ is obtained.

The boundary conditions at a TI can be derived in a dual fashion: let us consider a TI at which the dielectric permittivity or unit-length capacitance $\varepsilon$ is abruptly switched in time from $\varepsilon_1$ to $\varepsilon_2$, as sketched in Fig. 1c and 1d. Here we restrict ourselves to dielectric TIs, but similar arguments apply to the magnetic scenario. By integrating Ampere-Maxwell's law and Faraday's law [Eq. (1)] over a finite time interval, and letting this time-interval shrink to zero around the TI, we find that the integral of the magnetic field $H$ vanishes, the integral of the displacement field yields the difference $D_x(t_0^+) - D_x(t_0^-)$, and the integral of the electric current yields $\int_{t_0^-}^{t_0^+} J_s(t)dt = -\sigma_{se}$, representing the change in electric charge at the TI. Thus, the generalized temporal boundary conditions read

$$B_y(t_0^+) - B_y(t_0^-) = 0$$
$$D_x(t_0^+) - D_x(t_0^-) = \sigma_{se}, \quad (3)$$

where again $\sigma_{se}$ corresponds to charges acquired or lost at the TI, as a function of the process that drives the permittivity change in time. In the presence of a magnetic charge discontinuity, a term in the first equation of (3) may also arise.

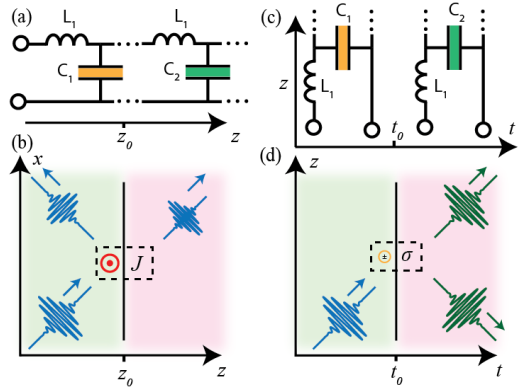

*Fig. 1:* (a,b) Spatial and (c,d) temporal interfaces, represented as (a,c) transmission lines with abruptly varying capacitance per unit length and (b,d) materials with abruptly varying permittivity. Dual to the way surface currents at a spatial interface may cause a discontinuity in the tangential magnetic field H, interface charges at a temporal interface can induce a discontinuity in the displacement field D.

In the case of charge conservation at the TI, we obtain the conventional boundary conditions, implying the continuity of $\mathbf{B}$ and $\mathbf{D}$, from which we can derive the well-established temporal scattering coefficients for the electric field[30,34-35]

$$T_D = \frac{\varepsilon_1}{2\varepsilon_2}\left(1 + \sqrt{\frac{\varepsilon_2}{\varepsilon_1}}\right); R_D = \frac{\varepsilon_1}{2\varepsilon_2}\left(1 - \sqrt{\frac{\varepsilon_2}{\varepsilon_1}}\right), \quad (4)$$

where $T$ is the temporal transmission (also known as forward wave) and $R$ is the temporal reflection (backward wave), and the subscripts $D$ denote the continuity of electric displacement.

The occurrence of 'surface charges' that arise at a TI may appear at first sight improbable, justifying the fact that they have been neglected in the broad literature on this topic. Yet, we show in the following that charge discontinuities naturally emerge in many instances, based on the microscopic implementation of TIs. Consider first a TI consisting of a sudden permittivity increase ($\varepsilon_2 > \varepsilon_1$), as sketched in Fig. 2a. In our transmission-line model, this may be obtained by adding a capacitor in the parallel branch of each L-C unit cell, which may be achieved through a switching element with dynamics much faster than the wave period. Indeed, this is the scheme used in the experimental demonstration of temporal reflections in[32], based on a loaded transmission-line metamaterial connected to a dense array of switched elements. After closing the switch, the total bound charge $Q$ per unit length originally present in the parallel branch of the transmission line is redistributed between the original unit cell capacitance and the additional one, while the voltage across the two parallel capacitors drops.

In the homogenization limit, consistent with Fig. 1d, this phenomenon corresponds to the conservation of the macroscopic displacement field $D$ and an abrupt drop in $E$, in agreement with the conventional conservation laws at a TI with $\varepsilon_2 > \varepsilon_1$ (assuming no dispersion). The corresponding scattering coefficients indeed read as in Eq. (4). The bottom panel of Fig. 2a shows the logarithm of the ratio between final and initial electromagnetic energy density $U \propto \varepsilon |E|^2 + \mu |H|^2$ for the forward (dashed line) and backward (dotted line) waves, their sum (total, continuous line) and the net electromagnetic (Minkowski) momentum density $\mathbf{P} = \mathbf{D} \times \mathbf{B}$ (corresponding to the difference between the forward and backward wave energy density, multiplied by $\sqrt{\varepsilon}$ ), as we vary the ratio $\varepsilon_2 / \varepsilon_1 > 1$. As expected, the time reflection grows as the permittivity contrast increases, and the total energy after the TI is reduced. This decrease in energy is consistent with the well-known two-capacitor problem[37], whereby the closing of a switch between a charged and an uncharged capacitor results in a net loss of energy via ohmic or radiative channels if the charge is conserved in the process. As expected, the total momentum $\mathbf{P}$ (red line) is conserved[33,36,38].

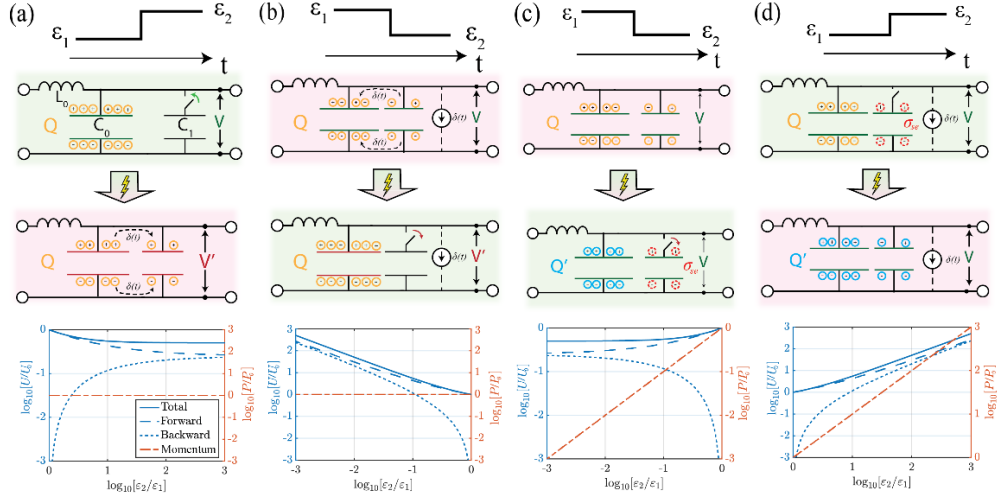

*Fig. 2:* (a-b) Assuming that the bound charge Q is conserved, a permittivity increase $\varepsilon_1 \to \varepsilon_2$ (a), represented in a transmission-line as the on-switching of a parallel lumped capacitor $C_1$, results in charge redistribution, which reduces the voltage V (electric field). In turn, this redistribution reduces (a, bottom) the total energy density U (continuous, dashed and dotted lines denote respectively total, forward and backward energy densities in logarithmic scale, normalized to the initial value $U_0$), while the momentum flux density (red dot-dashed line) is conserved ( $P_0$ denotes the initial momentum). By contrast, a (b) permittivity reduction can be achieved while conserving bound charges by transferring the charges before off-switching the capacitance $C_1$, which requires gain (realized, e.g., via a current source), thus (b, bottom) increasing the energy in the system. (c-d) Voltage (electric field) continuity instead is observed in a passive system upon a permittivity drop, whereby the bound charges in $C_1$ (depicted in red) are switched out of the system, (c, bottom) reducing the total energy, as well as the momentum density. Finally, imposing electric field continuity upon (d) a permittivity increase is equivalent to assuming that the additional capacitance enters the system with a dynamically varying, nonzero bound charge, produced, e.g., by a current source, in order to preserve the voltage, resulting in (d, bottom) an increase in total energy and momentum density.

The dual phenomenon corresponds to a reduction of permittivity (Fig. 2b), which may be obtained by switching off a capacitor $C_1$ initially present in the unit cell. However, in order for this scenario to conserve the total bound charge Q in the system, we cannot simply disconnect it from the transmission-line, but we need to make sure that a charge equal to the one present in $C_1$ at the instant of the TI is instantaneously transferred to the residual unit cell capacitance via an external mechanism (represented in the figure by a current generator). Given the need for a current generator in the presence of a finite voltage, we can expect the emergence of gain in this scenario, which is indeed consistent with the fact that the total energy

$U \sim D^2 / \varepsilon$ increases at a TI with $\varepsilon_2 < \varepsilon_1$ under the assumption that $D$ is conserved. Indeed, Fig. 2c (bottom) confirms that, as $\varepsilon_2 / \varepsilon_1 < 1$ decreases, the total energy in the system grows, together with a growing time-reflected wave. Also in this case the momentum is conserved in the process, since we conserve charge and $D$. A practical implementation of such a scenario is a TI in which the permittivity is reduced through a variable capacitor (varactor diode), which requires energy to decrease its capacitance in the presence of an applied voltage across its terminals conserving the charge.

The gain mechanism supporting the continuity of $D$ at a TI with $\varepsilon_2 < \varepsilon_1$ cannot be achieved through a switch, which was the mechanism used to generate TIs in Ref. 32. Indeed, in this experimental realization the TI was achieved by switching out the lumped capacitors previously loading the transmission-line, implying, as sketched in Fig. 2c, that charges may be lost upon switching (where "lost" is intended as charge being taken out of the medium supporting the wave propagation). This charge loss corresponds to a nonzero $\sigma_{se}$ in Eq. (3), consistent with the emergence of temporal surface charges at a TI that modify the boundary conditions and conservation laws. In this scenario, the displacement field is discontinuous, and the outflux or influx of charges at the TI needs to be considered. Since the two capacitors in Fig. 2c are in parallel, the off-switching of $C_1$ does not modify the voltage across the parallel branch, hence the $E$ field remains continuous at the TI. The resulting scattering coefficients for the electric field of the time-refracted and time-reflected waves become

$$T_E = \frac{1}{2}(1 + \sqrt{\frac{\varepsilon_1}{\varepsilon_2}}); R_E = \frac{1}{2}(1 - \sqrt{\frac{\varepsilon_1}{\varepsilon_2}}), \qquad (5)$$

which are consistent with those measured in 32 for TIs in which the permittivity decreases.

In the continuum limit, charge continuity implies that an instantaneous current source $J_s = -\partial_t[\varepsilon]\mathbf{E}$ emerges in Ampere-Maxwell's law

$$\nabla \times \mathbf{H} = \varepsilon \partial_t[\mathbf{E}] + \partial_t[\varepsilon]\mathbf{E} + \mathbf{J}_s, \qquad (6)$$

exactly canceling the term proportional to the time-derivative $\partial_t[\varepsilon]E_x$, and capturing the in/outflux of charges that leave (or enter) the system. As expected, both energy and momentum are reduced in this scenario (bottom), due to the loss of bound charges. This scenario corresponds to the sudden removal of polarizable matter, subtracting their electromagnetic energy and momentum from the system.

In a dual fashion, we may impose electric field continuity upon a TI involving a permittivity increase (Fig. 2d), by assuming that the additional capacitor $C_1$ enters the system with a nonzero initial distribution of bound charges, preventing charge redistribution. Here again $D$ is not continuous, implying the presence of the same source term in Maxwell's equations as the previous case, but here with the effect of adding bound charges to the system, increasing energy and momentum (bottom). This scenario is equivalent to adding polarizable matter, which contributes its initial energy and momentum.

While the four examples discussed in Fig. 2 correspond in pairs to the same *macroscopic* scattering phenomenon (time-interfaces with increasing or decreasing permittivity), the underlying *microscopic* implementation substantially changes the conservation laws, the boundary conditions, and the overall electromagnetic result. SM Sec. 1 shows time-domain numerical solutions for the respective scenarios, demonstrating the importance of carefully selecting the appropriate boundary conditions in the continuous modeling of TIs that describe their actual implementation.

We stress that these results affect both sub-cycle switching processes, described by TIs and associated boundary conditions, as well as continuous parametric processes, in which again the continuity of various quantities along the process needs to be carefully assessed based on the underlying mechanisms and phenomena. Although so far we have highlighted the importance of the microscopic implementation of TIs via a circuit picture, these electrodynamic considerations are very much relevant also in the context of ultrafast optical switching in natural materials. For instance, the effect of a pump beam in a nonlinear material may conserve bound charges and decrease the effective permittivity by stretching lattice bonds in a phononic crystal 39, or in turn it may produce interface charges by creating or recombining electron-hole pairs in a semiconductor, resulting in different conservation laws and macroscopic boundary conditions, which need to be fully considered to be able to predict the electromagnetic scattering at such TIs.

Although Fig. 2 considers two extreme scenarios in which $D$ or $E$ are continuous, real systems may involve mixed boundary conditions, which conserve only a fraction of the bound charge in the system, or other quantities, since different microscopic phenomena may coexist in the specific realization of a TI. In general, an arbitrary interface charge $\sigma_{se}$ in the displacement field yields generalized scattering coefficients

$$T_{\sigma_{se}} = \frac{\varepsilon_1}{2\varepsilon_2}\left(1 + \sqrt{\frac{\varepsilon_2}{\varepsilon_1}} + \bar{\sigma}_{se}\right)$$
$$R_{\sigma_{se}} = \frac{\varepsilon_1}{2\varepsilon_2}\left(1 - \sqrt{\frac{\varepsilon_2}{\varepsilon_1}} + \bar{\sigma}_{se}\right)$$

(7)

where $\bar{\sigma}_{se} = \sigma_{se}/D_x(t^-)$ is the ratio between the interface and bound charges at the TI. Note how both forward and backward waves are equally affected by the charge term, meaning that such charge discontinuity cannot redirect power, but only damp or enhance both forward and backward waves. Interestingly, there is a condition on the interface charge $\sigma_{se}$ whereby time-reflected waves are entirely eliminated, a form of impedance matching for the time-interface. This occurs concurrently with the exact matching between the energy lost by the partial charge sink and the gain deriving from the permittivity reduction upon partial charge conservation, as discussed in Sec. 2 of the SM. By contrast, maximum charge discontinuity $\bar{\sigma}_{se}$ is obtained in the limit of totally disconnecting the additional capacitance, i.e., $\bar{\sigma}_{se,max} = (\varepsilon_2 - \varepsilon_1)/\varepsilon_1$, which corresponds to the case of electric field continuity. In SM Sec. 3, we derive generalized formulas for the electromagnetic energy and momentum density under arbitrary charge discontinuity.

These findings may also be relevant in the context of the ongoing discussion on the level of energy required to drive time crystals and space-time metamaterials[40-42]. In these discussions, it is clear that proper boundary conditions and specific microscopic phenomena need to be carefully considered to account for the emergent macroscopic dynamics. The short timescales involved in ultrafast processes involve by definition large bandwidths and, especially when considering polaritonic materials with large nonlinearities, it becomes necessary to account for the effect of temporal dispersion. In this context, our findings can be straightforwardly extended to dispersive media, for which the presence of interface charges may have other interesting implications. As a canonical example, consider a Drude medium in which the plasma frequency is abruptly changed in time. This phenomenon may arise upon a modification of carrier density, e.g., through interband pumping, or of effective mass, if carriers are driven towards a non-parabolic regime. These opportunities connect to the recent interest in polaritonic materials for time-interfaces, as in the case of indium tin oxide 43 whose ultrafast nonlinearities are being discussed as an excellent platform for observing these phenomena in optics[44-46]. Switching the plasma frequency has been also studied by the plasma physics community (see, e.g., Refs 34-35,47-48), typically assuming the continuity of current density $\mathbf{J}_c$, electric field $\mathbf{E}$ and magnetic field $\mathbf{H}$ [35], consistent with the modelling of continuous temporal variations of Drude media[28-29,49-50]. Next, we explore the impact of different microscopic implementations on TIs involving dispersive media.

To account for the Drude response, we introduce an additional equation for the Drude current $\mathbf{J}_c$ in Eq. (1). Starting from the equation of motion for a single free electron $\partial_t[m\partial_t x] = \partial_t m \partial_t x + m\partial_t^2 x = -eE_x$, we define the macroscopic current $J_c = -eN\partial_t x$, with the time-dependent density of electrons $N(t)$ and effective mass $m(t)$, and include a voltage (electric field) source term $E_s$ to obtain the general equation

$$\partial_t[(\frac{m}{e^2N})J_c] = \partial_t\varphi = L\partial_t[J_c] + \partial_t[L]J_c = E_x + E_s. \tag{8}$$

The expression $m/(e^2N) \equiv L(t) = [\varepsilon_0\omega_p^2(t)]^{-1}$ defines the inductance of the Drude electrons, while the product $LJ_c = \varphi$ corresponds to the magnetic flux linkage resulting from the conduction current, which behaves as the electromagnetic momentum contributed by the Drude electrons. Importantly, the time derivative of $L$ in Eq. (8), which reduces to a delta function for an instantaneous switching event, contributes part of the magnetic flux linkage entering or leaving the system, parallel to the term $\partial_t[\varepsilon]$ in Eq. (6), and it is similarly dependent on the microscopic implementation of a TI.

Fig. 3 shows a transmission-line representation of a time-interface involving a Drude material, analogous to Fig. 2 for the non-dispersive case. Here, an inductive shunt branch describes the Drude electron contribution. Assuming first that the plasma frequency is increased at a TI (Fig. 3a), we can model this effect by shorting one of two series inductors in the Drude inductive branch, where the total flux linkage in the shunt branch is $\varphi = (L_a + L_b)J_c = L_sJ_c$ before the TI. At the instant when $L_b$ is shorted, the magnetic flux linkage $L_bJ_c$ stored in it is lost, similar to when we switched off the electric charges in $C_1$ in Fig. 2c. In a mechanical analogy, this phenomenon is equivalent to dropping a portion of the mass while it is moving at a finite velocity, such that some of the momentum in the system is abruptly lost. In this scenario, the current $J_c$, describing the velocity of all electrons before a fraction of them leaves the system, must clearly be continuous, along with the electric field $E_x$, i.e., the voltage across the shunt branches, and with the magnetic field, i.e., the current along the transmission-line.

These boundary conditions are consistent with what is generally assumed in the literature on time-interfaces in plasmas or dispersive media[35]. The more general model shown in Eq. (8), however, highlights how the assumption of current continuity implies that the term proportional to the derivative of the inductance $\partial_t[L]J_c$ on the left side of Eq. (8) must be compensated by a corresponding flux linkage sink term to account for the energy stored in $L_b$ just before the TI, which is dissipated by its internal resistance upon shorting it. As demonstrated explicitly in[35], these boundary conditions result in the scattering coefficients

$$T_{J_c} = \frac{\omega_1(\omega_2 + \omega_1)}{2\omega_2^2}; R_{J_c} = \frac{\omega_1(\omega_2 - \omega_1)}{2\omega_2^2}; W_{J_c} = \frac{\omega_{p,2}^2}{\omega_2^2}(1 - \omega_{p,1}^2/\omega_{p,2}^2) \tag{9}$$

for the magnetic field, corresponding to forward traveling waves, backward traveling waves, and a DC mode comprising only conduction current and magnetic field, also called the "wiggler mode"[35], where $\omega_j = \sqrt{(\omega_{p,j}^2 + \omega_0^2)}$, $\omega_0 = c_0k$, and the subscript $J_c$ indicates that the conduction current is conserved at the temporal boundary.

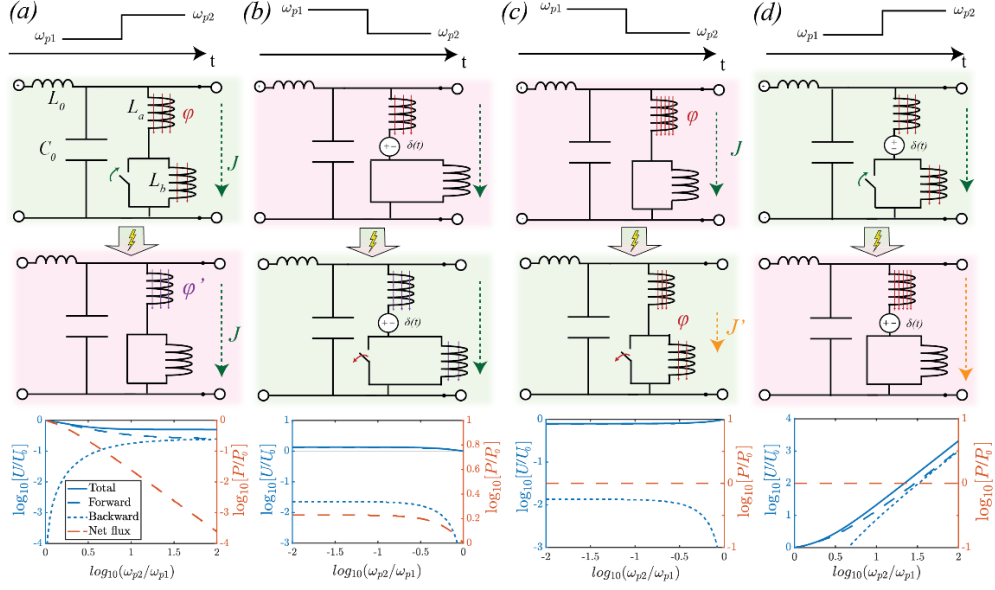

***Figure 3:*** *(a, top) An increase in plasma frequency can be modeled as the shorting of a series inductor in the shunt branch of a transmission-line. In this scenario, the current $J$ in the branch remains unchanged, along with the flux linkage on $L_a$, whereas the flux on $L_b$ is lost. Mechanically, this scenario is analogous to particles leaving the system (e.g. droplets leaving a water bucket) with their instantaneous velocity, carrying their momentum away. The analogy is corroborated by panel (a, bottom), which shows how the total energy density (continuous blue line) in the system is reduced, along with the energy in the forward (dashed line), and the electromagnetic momentum $P$ (dashed red line) is reduced. By contrast, (b, top) reducing the plasma frequency by opening the short can only preserve the current in the branch if the additional inductor $L_b$ is prepared in such a state that its current at the switching instant matches the one of the original inductor, (c, bottom) increasing the total energy and net momentum density. This is analogous to additional particles entering a mechanical system with its identical instantaneous velocity, thereby increasing its total momentum without affecting its velocity. Instead, in a passive system, (d, top) voltage spikes on the respective inductors resulting from the TI drive a finite current through the additional inductor $L_b$, while abruptly reducing the current on $L_a$ within an infinitesimal time interval, such that the total flux linkage is redistributed between the two inductors, reducing the current $J_c$ across the branch. Mechanically, this is analogous to adding particles with no initial velocity into the system, and letting the system redistribute its momentum among them. As a result, (c, bottom) the total electromagnetic momentum is conserved, whereas the total energy is reduced. (d, top) In a dual fashion, conserving flux linkage upon an increase in plasma frequency would require a gain mechanism capable of replacing the flux lost by shorting the inductor $L_b$ in the branch by increasing the current on $L_a$. Expectedly, (d, bottom) this increases the energy density in the system, while conserving its momentum.*

Fig. 3a (bottom) shows the energy density

$$U \propto [(\varepsilon + \partial_\omega \varepsilon)|E|^2 + \mu |H|^2] \tag{10}$$

and net momentum

$$P \propto [(\varepsilon + \omega \partial_\omega \varepsilon)|E|^2 + \mu |H|^2]/\omega, \tag{11}$$

for propagating waves in a dispersive medium[51] before and after a TI (where $\varepsilon(\omega) = 1 - \omega_p^2/\omega^2$ and $\mu = \mu_0 = 1$), demonstrating that the energy density is reduced along with part of the electromagnetic momentum (dashed red line), as expected from the loss of flux linkage.

Let us now consider the opposite scenario, whereby the second inductor is initially shorted and then connected to the shunt branch, as in Fig. 3b. To avoid a drop in current density resulting from the introduction of the additional inductor $L_b$, the latter must be prepared with an initial current $J_b = J_c$, identical to the one initially on $L_a$, upon entering the system. Clearly, the total flux linkage here increases from $\varphi = L_a J_c$ to

$\varphi' = (L_a + L_b)J_c$, together with the total energy density and net momentum in the system (Fig. 3b, bottom). This is analogous to introducing additional particles in a mechanical system at the same velocity as the initial ones, increasing the total momentum of the system. This is not surprising, since a time-interface conserving current upon a plasma frequency reduction must necessarily provide gain. In fact, both the reduction and increase in flux linkage $LJ_c$, which guarantee current conservation, occur via an external voltage sink or source $E_s$, which acts as a $\delta$-function at the instant of switching, dual to the mechanism that removes or introduces extra charges in the active, non-dispersive scenarios, cancelling the term $\partial_t[L]J_c$ in Eq. (8).

Fig. 3c considers the scenario in which we open the short without providing an external kick, i.e., in a passive fashion. As the switch is opened, a voltage spike, dual to the current spike that occurs upon closing the switch in the non-dispersive case (see Fig. 2a), stabilizes the current in the shunt inductive branch by forcing a current $J_c' = (L_1/L_s)J_c$ through $L_2$, which ensures that the flux linkage $\varphi$ is conserved. Here it is important to point out that an instantaneous model of this problem is only appropriate when the microscopic dynamics is much faster than the field oscillations. This is analogous to a mechanical scenario where additional particles are added to the system with zero initial velocity, thus conserving the total momentum of the system, which is instead redistributed among them, such that the particle velocity is instead reduced, in analogy with the current $J_c \to J_c'$. This condition corresponds precisely to the case where the additional term in Eq. (8), which features the time-derivative of the inductance, is not counteracted by a source, and therefore acts as a current sink. Applying accordingly the continuity of $\mathbf{E}$, $\mathbf{H}$ and $\varphi$ yields magnetic-field scattering coefficients (see SM Sec. 4)

$$T_\varphi = \frac{\omega_2 + \omega_1}{2\omega_1}; R_\varphi = \frac{\omega_2 - \omega_1}{2\omega_1}; W_\varphi = 0, \qquad (12)$$

where the subscript $\varphi$ denotes flux linkage conservation. Importantly, (Fig. 3c, bottom) conservation of the net momentum density cements our mechanical analogy, while the total energy density is reduced as expected.

By contrast, shorting the additional inductor $L_b$ while simultaneously applying an external instantaneous voltage across $L_a$ via an auxiliary source (Fig. 5d) can compensate for the flux linkage $L_bJ_c$ lost by shorting $L_b$, changing the current in $L_a$ from $J_c$ to $J_c' = [(L_a + L_b)/L_a]J_c$. This dynamic is equivalent to accelerating the remaining particles in a system to match the total momentum it carried before a fraction of them were removed, and generally results in (bottom panel) an increase of the total energy in the system. Note how, under this boundary condition, the amplitude of the DC wiggler mode is identically zero. In Fig. 3c, this is because there is no flux left in $L_b$ to form a local DC current, which is present instead in e.g. panel (a). This is also the case for panel (d), where the additional voltage source provides the exact amount of flux lost upon shorting $L_b$, thus cancelling any hanging DC current. Plots of the scattering coefficients for the respective cases are given in SM Sec. 5, highlighting widely different scattering phenomena emerging as a function of the underlying boundary conditions and conservation laws, while numerical field plots are provided in SM Sec. 6.

Finally, an interesting duality underpins the phenomenology of TIs in dispersive and non-dispersive media: the loss naturally occurring upon charge redistribution in the non-dispersive scenario in Fig. 2a is in fact dual to the one occurring upon flux redistribution in the dispersive scenario of Fig. 3c. Similarly, the loss of charge into a decoupled DC mode in the non-dispersive scenario shown in Fig. 2c parallels the loss of flux into the DC/wiggler mode of Fig. 3a. Identical parallels exist between Fig. 2b and Fig. 3d, whereby charge and flux are respectively pumped into one of the two capacitors or inductors by an external pumping mechanism, and to Fig. 2d and Fig. 3b, whereby the additional capacitors or inductors entering the system are prepared in an initial state which matches respectively the voltage or current just before the TI. In general, our framework identifies how flux linkage in the dispersive scenario mimics the role of bound charges at a non-dispersive TI.

**Discussion**

In this work, we have demonstrated the paramount importance of considering the microscopic implementation of time-interfaces to properly model their rich electrodynamics. Boundary conditions and associated conservation laws at temporal scattering phenomena can be profoundly affected by the specific mechanisms underlying time variations in materials, offering opportunities for rich physics discoveries. Our results have implications for both quasi-instantaneous switching events, as well as continuous modulations, and overcome long-held misconceptions on the physics of time-varying photonic systems by exposing problems and opportunities that arise from the use of proper boundary conditions, both to model existing and future experiments[21,27-29,32,49] and to explore new regimes for time-interfaces, time-crystals and space-time metamaterials. We considered both non-dispersive and dispersive materials, paving the way towards a wider exploration of the physics behind time-modulated and time-switched photonic systems, which may generally include structures characterized by additional modes, stemming from geometric or material resonances, e.g., in phononic systems and photonic crystals. Finally, amidst the current exploration of ultrafast femtosecond-timescale responses exhibited by transparent conducting oxides in the near-infrared[2-3,27-29], our work opens new vistas towards the macroscopic modelling of ultrafast nonlinearities, complementing the microscopic nonequilibrium dynamics currently under intense scrutiny[44-46].

**Funding.** Simons Foundation (733684, 855344 and Collaboration on Extreme Wave Phenomena, AA, NE, EG); Spanish Ministry of Universities under a María Zambrano Grant (DMS).

**Disclosures**. The authors declare no conflicts of interest.

**Data availability**. All data shown in this article are freely available from the Authors upon reasonable request.

**Supplementary material**. Supplementary material is available in the online version of this article.

**References**

1. Pendry, J. B. (2008). Time reversal and negative refraction. *Science*, *322*(5898), 71-73.
2. Bruno, V., DeVault, C., Vezzoli, S., et al. (2020). Negative refraction in time-varying strongly coupled plasmonic-antenna–epsilon-near-zero systems. *Physical Review Letters*, *124*(4), 043902.
3. Prain, A., Vezzoli, S., Westerberg, N., et al. (2017). Spontaneous photon production in time-dependent epsilon-near-zero materials. *Physical Review Letters*, *118*(13), 133904.
4. Othman, M. A., & Capolino, F. (2017). Theory of exceptional points of degeneracy in uniform coupled waveguides and balance of gain and loss. *IEEE Transactions on Antennas and Propagation*, *65*(10), 5289-5302.
5. Wang, X., Mirmoosa, M. S., Asadchy, V. S., et al. (2023). Metasurface-based realization of photonic time crystals. *Science Advances*, *9*(14), eadg7541.
6. Nation, P. D., Johansson, J. R., Blencowe, M. P., et al. (2012). Colloquium: Stimulating uncertainty: Amplifying the quantum vacuum with superconducting circuits. *Reviews of Modern Physics*, *84*(1), 1.
7. Galiffi, E., Huidobro, P. A., & Pendry, J. B. (2019). Broadband nonreciprocal amplification in luminal metamaterials. *Physical Review Letters*, *123*(20), 206101.
8. Horsley, S. A., & Pendry, J. B. (2023). Quantum electrodynamics of time-varying gratings. *Proceedings of the National Academy of Sciences*, *120*(36), e2302652120.
9. Cheng, D., Wang, K., & Fan, S. (2023). Artificial non-Abelian lattice gauge fields for photons in the synthetic frequency dimension. *Physical Review Letters*, *130*(8), 083601.
10. Lustig, E., Sharabi, Y., & Segev, M. (2018). Topological aspects of photonic time crystals. *Optica*, *5*(11), 1390-1395.
11. Huidobro, P. A., Galiffi, E., Guenneau, S., et al. (2019). Fresnel drag in space–time-modulated metamaterials. *Proceedings of the National Academy of Sciences*, *116*(50), 24943-24948.
12. Yu, Z., & Fan, S. (2009). Complete optical isolation created by indirect interband photonic transitions. *Nature Photonics*, *3*(2), 91-94.
13. Fleury, R., Sounas, D. L., Sieck, C. F., et al. (2014). Sound isolation and giant linear nonreciprocity in a compact acoustic circulator. *Science*, *343*(6170), 516-519.
14. Sounas, D. L., & Alù, A. (2017). Non-reciprocal photonics based on time modulation. *Nature Photonics*, *11*(12), 774-783.

15. Tzuang, L. D., Fang, K., Nussenzveig, P., et al. (2014). Non-reciprocal phase shift induced by an effective magnetic flux for light. *Nature Photonics*, *8*(9), 701-705.
16. Deck-Léger, Z. L., Chamanara, N., Skorobogatiy, M., et al. (2019). Uniform-velocity spacetime crystals. *Advanced Photonics*, *1*(5), 056002.
17. Camacho, M., Edwards, B., & Engheta, N. (2020). Achieving asymmetry and trapping in diffusion with spatiotemporal metamaterials. *Nature Communications*, *11*(1), 3733.
18. Buddhiraju, S., Li, W., & Fan, S. (2020). Photonic refrigeration from time-modulated thermal emission. *Physical Review Letters*, *124*(7), 077402.
19. Vázquez-Lozano, J. E., & Liberal, I. (2023). Incandescent temporal metamaterials. *Nature Communications*, *14*(1), 4606.
20. Yu, R., & Fan, S. (2023). Manipulating coherence of near-field thermal radiation in time-modulated systems. *Physical Review Letters*, *130*(9), 096902.
21. Galiffi, E., Xu, G., Yin, S., et al. (2023). Broadband coherent wave control through photonic collisions at time interfaces. *Nature Physics*, *19*(11), 1703-1708.
22. Lyubarov, M., Lumer, Y., Dikopoltsev, A., et al. (2022). Amplified emission and lasing in photonic time crystals. *Science*, *377*(6604), 425-428.
23. Pacheco-Peña, V., & Engheta, N. (2020). Antireflection temporal coatings. *Optica*, *7*(4), 323-331.
24. Liberal, I., Vázquez-Lozano, J. E., & Pacheco-Peña, V. (2023). Quantum antireflection temporal coatings: quantum state frequency shifting and inhibited thermal noise amplification. *Laser & Photonics Reviews*, *17*(9), 2200720.
25. Pacheco-Peña, V., & Engheta, N. (2020). Temporal aiming. *Light: Science & Applications*, *9*(1), 129.
26. Galiffi, E., Tirole, R., Yin, S., et al. (2022). Photonics of time-varying media. *Advanced Photonics*, *4*(1), 014002.
27. Zhou, Y., Alam, M. Z., Karimi, M., et al. (2020). Broadband frequency translation through time refraction in an epsilon-near-zero material. *Nature Communications*, *11*(1), 2180.
28. Lustig, E., Segal, O., Saha, S., et al. (2023). Time-refraction optics with single cycle modulation. *Nanophotonics*, *12*(12), 2221-2230.
29. Tirole, R., Vezzoli, S., Galiffi, E., et al. (2023). Double-slit time diffraction at optical frequencies. *Nature Physics*, *19*, 999-1002.
30. Morgenthaler, F. R. (1958). Velocity modulation of electromagnetic waves. *IRE Transactions on microwave theory and techniques*, *6*(2), 167-172.
31. Bacot, V., Labousse, M., Eddi, A., et al. (2016). Time reversal and holography with spacetime transformations. *Nature Physics*, *12*(10), 972-977.
32. Moussa, H., Xu, G., Yin, S., et al. (2023). Observation of temporal reflection and broadband frequency translation at photonic time interfaces. *Nature Physics*, *19*, 863-868.
33. Galiffi, E., Yin, S., & Alú, A. (2022). Tapered photonic switching. *Nanophotonics*, *11*(16), 3575-3581.
34. Mendonça, J. T., & Shukla, P. K. (2002). Time refraction and time reflection: two basic concepts. *Physica Scripta*, *65*(2), 160.
35. Kalluri, D. K. (2018). Electromagnetics of time varying complex media: frequency and polarization transformer. CRC Press. *Boca Raton*.
36. Kong, J. A. (1975). Theory of electromagnetic waves. EMW Publishing. *New York*.
37. Zucker, C. (1955). Condenser problem. *American Journal of Physics*, *23*(7), 469-469.
38. Ortega-Gomez, A., Lobet, M., Vázquez-Lozano, J. E., et al. (2023). Tutorial on the conservation of momentum in photonic time-varying media. *Optical Materials Express*, *13*(6), 1598-1608.
39. Cartella, A., Nova, T. F., Fechner, M., et al. (2018). Parametric amplification of optical phonons. *Proceedings of the National Academy of Sciences*, *115*(48), 12148-12151.
40. Hayran, Z., Khurgin, J. B., & Monticone, F. (2022). ℏω versus ℏk: dispersion and energy constraints on time-varying photonic materials and time crystals. *Optical Materials Express*, *12*(10), 3904-3917.
41. Khurgin, J. B. (2023). Energy and power requirements for alteration of the refractive index. *Laser & Photonics Reviews*, 18(4), 2300836.
42. Khurgin, J. B. (2023). Photonic time crystals and parametric amplification: similarity and distinction. *arXiv preprint arXiv:2305.15243*.
43. Lobet, M., Kinsey, N., Liberal, I., et al. (2023). New horizons in near-zero refractive index photonics and hyperbolic metamaterials. *ACS photonics,* 10(11), 3805-3820.
44. Narimanov, E. E. (2023). Ultrafast Optical Modulation by Virtual Interband Transitions. *arXiv preprint arXiv:2310.15908*.
45. Un, I. W., Sarkar, S., & Sivan, Y. (2023). Electronic-Based Model of the Optical Nonlinearity of Low-Electron-Density Drude Materials. *Physical Review Applied*, *19*(4), 044043.
46. Sarkar, S., Un, I. W., & Sivan, Y. (2023). Electronic and Thermal Response of Low-Electron-Density Drude Materials to Ultrafast Optical Illumination. *Physical Review Applied*, *19*(1), 014005.
47. Nishida, A., Yugami, N., Higashiguchi, T., et al. (2012). Experimental observation of frequency up-conversion by flash ionization. *Applied Physics Letters*, 101(16), 161118.
48. Avitzour, Y., & Shvets, G. (2008). Manipulating electromagnetic waves in magnetized plasmas: Compression, frequency shifting, and release. *Physical Review Letters*, 100(6), 065006.
49. Bohn, J., Luk, T. S., Tollerton, C., et al. (2021). All-optical switching of an epsilon-near-zero plasmon resonance in indium tin oxide. *Nature communications*, *12*(1), 1017.

50. Horsley, S. A. R., Galiffi, E., & Wang, Y. T. (2023). Eigenpulses of dispersive time-varying media. *Physical Review Letters*, *130*(20), 203803.
51. Bliokh, K. Y., Bekshaev, A. Y., & Nori, F. (2017). Optical momentum, spin, and angular momentum in dispersive media. *Physical Review Letters*, *119*(7), 073901.